\documentclass[prd,aps,showpacs,epsf,floats,10pt]{revtex4}
\usepackage{amssymb}
\usepackage{amsfonts}
\usepackage{amsmath}

\input{tcilatex}

\begin{document}

\title{Recent theoretical progress on an information geometrodynamical
approach to chaos}
\author{Carlo Cafaro}
\email{carlocafaro2000@yahoo.it}
\affiliation{Department of Physics, University at Albany-SUNY,1400 Washington Avenue,
Albany, NY 12222, USA}

\begin{abstract}
In this paper, we report our latest research on a novel theoretical
information-geometric framework suitable to characterize chaotic dynamical
behavior of arbitrary complex systems on curved statistical manifolds.
Specifically, an information-geometric analogue of the Zurek-Paz quantum
chaos criterion of linear entropy growth and an information-geometric
characterization of chaotic (integrable) energy level statistics of a
quantum antiferromagnetic Ising spin chain in a tilted (transverse) external
magnetic field are presented.
\end{abstract}

\maketitle


\section{\textbf{Introduction}}

Research on complexity \cite{gell-mann, crutchfield} has created a new set
of ideas on how very simple systems may give rise to very complex behaviors.
Moreover, in many cases, the "laws of complexity" have been found to hold
universally, caring not at all for the details of the system's constituents.
Chaotic behavior is a particular case of complex behavior and it will be the
object of the present work.

In this paper we make use of the so-called Entropic Dynamics (ED) \cite%
{caticha1}. ED is a theoretical framework that arises from the combination
of inductive inference (Maximum Entropy Methods (ME), \cite{caticha1A,
caticha2}) and Information Geometry (IG) \cite{amari}. The most intriguing
question being pursued in ED stems from the possibility of deriving dynamics
from purely entropic arguments. This is clearly valuable in circumstances
where microscopic dynamics may be too far removed from the phenomena of
interest, such as in complex biological or ecological systems, or where it
may just be unknown or perhaps even nonexistent, as in economics. It has
already been shown that entropic arguments do account for a substantial part
of the formalism of quantum mechanics, a theory that is presumably
fundamental \cite{caticha3}. Laws of physics may just be consistent,
objective ways to manipulate information. Following this line of thought, we
extend the applicability of ED to temporally-complex (chaotic) dynamical
systems on curved statistical manifolds and identify relevant measures of
chaoticity of such an information geometrodynamical approach to chaos (IGAC).

The layout of the paper is as follows. In the section II, we give an
introduction to the main features of our IGAC. In section III, we apply our
theoretical construct to three complex systems. We emphasize that we have
omitted technical details some of which can be found in our previous
articles \cite{cafaro1, cafaro2, cafaro3, cafaro4, cafaro5, cafaro6,
cafaro7, cafaro8}. Finally, in section IV we present our final remarks and
future research directions.

\section{General Formalism of the IGAC}

The IGAC is an application of ED to complex systems of arbitrary nature. ED
is a form of information-constrained dynamics built on curved statistical
manifolds $\mathcal{M}_{S}$ where elements of the manifold are probability
distributions $\left\{ P\left( X|\Theta \right) \right\} $ that are in a
one-to-one relation with a suitable set of macroscopic statistical variables 
$\left\{ \Theta \right\} $ that provide a convenient parametrization of
points on $\mathcal{M}_{S}$. The set $\left\{ \Theta \right\} $ is called
the \textit{parameter space }$\mathcal{D}_{\Theta }$ of the system.

In what follows, we schematically outline the main points underlying the
construction of an arbitrary form of entropic dynamics. First, the
microstates of the system under investigation must be defined. For the sake
of simplicity, we assume our system is characterized by an $l$-dimensional 
\textit{microspace} with microstates $\left\{ x_{i}\right\} $ where $i=1$%
,..., $l$. The main goal of an ED model is that of inferring "macroscopic
predictions" in the absence of detailed knowledge of the microscopic nature
of arbitrary complex systems. Once the microstates have been defined, we
have to select the relevant information about such microstates. In other
words, we have to select the \textit{macrospace} of the system. For the sake
of the argument, we assume that our microstates are Gaussian-distributed.
They are defined by $2l$-information constraints, for example their
expectation values $\mu _{i}$ and variances $\sigma _{i}$. In addition to
information constraints, each Gaussian distribution $p_{k}\left( x_{k}|\mu
_{k}\text{, }\sigma _{k}\right) $ of each microstate $x_{k}$ must satisfy
the usual normalization conditions. Once the microstates have been defined
and the relevant information constraints selected, we are left with a set of
probability distributions $p\left( X|\Theta \right) =\underset{k=1}{\overset{%
l}{\dprod }}$ $p_{k}\left( x_{k}|\mu _{k}\text{, }\sigma _{k}\right) $
encoding the relevant available information about the system where $X$ is
the $l$-dimensional microscopic vector with components $\left( x_{1}\text{%
,...,}x_{l}\right) $ and $\Theta $ is the $2l$-dimensional macroscopic
vector with coordinates $\left( \mu _{1}\text{,..., }\mu _{l}\text{; }\sigma
_{1}\text{,..., }\sigma _{l}\right) $. The set $\left\{ \Theta \right\} $
define the $2l$-dimensional space of macrostates of the system, the
statistical manifold $\mathcal{M}_{S}$. A measure of distinguishability
among macrostates is obtained by assigning a probability distribution $%
P\left( X|\Theta \right) \ni \mathcal{M}_{S}$ to each macrostate $\Theta $ .
Assignment of a probability distribution to each state endows $\mathcal{M}%
_{S}$ with a metric structure. Specifically, the Fisher-Rao information
metric $g_{\mu \nu }\left( \Theta \right) $ \cite{amari} defines a measure
of distinguishability among macrostates on $\mathcal{M}_{S}$, with $\mathcal{%
M}_{S}$ being defined as the set of probabilities $\left\{ p\left( X|\Theta
\right) \right\} $ described above where $X\in 
\mathbb{R}
^{3N}$, $\Theta \in \mathcal{D}_{\Theta }=\left[ \mathcal{I}_{\mu }\times 
\mathcal{I}_{\sigma }\right] ^{3N}$. The parameter space $\mathcal{D}%
_{\Theta }$ (homeomorphic to $\mathcal{M}_{S}$) is the direct product of the
parameter subspaces $\mathcal{I}_{\mu }$ and $\mathcal{I}_{\sigma }$, where $%
\mathcal{I}_{\mu }=\left( -\infty \text{, }+\infty \right) _{\mu }$ and $%
\mathcal{I}_{\sigma }=\left( 0\text{, }+\infty \right) _{\sigma }$ in the
conventional Gaussian case. Once $\mathcal{M}_{S}$ and $\mathcal{D}_{\Theta
} $ are defined, the ED formalism provides the tools to explore dynamics
driven \ on $\mathcal{M}_{S}$\ by entropic arguments. Specifically, given a
known initial macrostate $\Theta ^{\left( \text{initial}\right) }$
(probability distribution), and that the system evolves to a final known
macrostate $\Theta ^{\left( \text{final}\right) }$, the possible
trajectories of the system are examined in the ED approach using ME methods.

The geodesic equations for the macrovariables of the Gaussian ED model are
given by\textit{\ nonlinear} second order coupled ordinary differential
equations. They describe a \textit{reversible} dynamics whose solution is
the trajectory between an initial $\Theta ^{\left( \text{initial}\right) }$
and a final macrostate $\Theta ^{\left( \text{final}\right) }$. Given the
Fisher-Rao information metric, we can apply standard methods of Riemannian
differential geometry to study the information-geometric structure of the
manifold $\mathcal{M}_{S}$ underlying the entropic dynamics. Connection
coefficients $\Gamma _{\mu \nu }^{\rho }$, Ricci tensor $R_{\mu \nu }$,
Riemannian curvature tensor $R_{\mu \nu \rho \sigma }$, sectional curvatures 
$\mathcal{K}_{\mathcal{M}_{S}}$, scalar curvature $\mathcal{R}_{\mathcal{M}%
_{S}}$, Weyl anisotropy tensor $W_{\mu \nu \rho \sigma }$, Killing fields $%
\xi ^{\mu }$ and Jacobi fields $J^{\mu }$ can be calculated in the usual way.

To characterize the chaotic behavior of complex entropic dynamical systems,
we are mainly concerned with the signs of the scalar and sectional
curvatures of $\mathcal{M}_{S}$, the asymptotic behavior of Jacobi fields $%
J^{\mu }$ on $\mathcal{M}_{S}$, the existence of Killing vectors $\xi ^{\mu
} $ and the asymptotic behavior of the information-geometrodynamical entropy
(IGE) $\mathcal{S}_{\mathcal{M}_{S}}$ (see (\ref{IGE})). It is crucial to
observe that true chaos is identified by the occurrence of two features: 1)
strong dependence on initial conditions and exponential divergence of the
Jacobi vector field intensity, i.e., \textit{stretching} of dynamical
trajectories; 2) compactness of the configuration space manifold, i.e., 
\textit{folding} of dynamical trajectories. The negativity of the Ricci
scalar $\mathcal{R}_{\mathcal{M}_{S}}$ implies the existence of expanding
directions in the configuration space manifold $\mathcal{M}_{s}$. Indeed,
since $\mathcal{R}_{\mathcal{M}_{S}}$ is the sum of all sectional curvatures
of planes spanned by pairs of orthonormal basis elements $\left\{ e_{\rho
}=\partial _{\Theta _{\rho }}\right\} $, the negativity of the Ricci scalar
is only a \textit{sufficient} (not necessary) condition for local
instability of geodesic flow. For this reason, the negativity of the scalar
provides a \textit{strong }criterion of local instability. A powerful
mathematical tool we use to investigate the stability or instability of a
geodesic flow is the Jacobi-Levi-Civita equation (JLC-equation) for geodesic
spread. Finally, the asymptotic regime of diffusive evolution describing the
possible exponential increase of average volume elements on $\mathcal{M}_{s}$
provides another useful indicator of dynamical chaoticity. The exponential
instability characteristic of chaos forces the system to rapidly explore
large areas (volumes) of the statistical manifold. It is interesting to note
that this asymptotic behavior appears also in the conventional description
of quantum chaos where the von Neumann entropy increases linearly at a rate
determined by the Lyapunov exponents. The linear increase of entropy as a
quantum chaos criterion was introduced by Zurek and Paz \cite{zurek1,
zurek1A}. In our information-geometric approach a relevant quantity that can
be useful to study the degree of instability characterizing ED models is the
information geometrodynamical entropy (IGE) defined as \cite{cafaro2},%
\begin{equation}
\mathcal{S}_{\mathcal{M}_{s}}\left( \tau \right) \overset{\text{def}}{=}%
\underset{\tau \rightarrow \infty }{\lim }\log \mathcal{V}_{\mathcal{M}_{s}}%
\text{ with }\mathcal{V}_{\mathcal{M}_{s}}\left( \tau \right) =\frac{1}{\tau 
}\dint\limits_{0}^{\tau }d\tau ^{\prime }\left( \underset{\mathcal{M}%
_{s}\left( \tau ^{\prime }\right) }{\int }\sqrt{g}d^{2l}\Theta \right)
\label{IGE}
\end{equation}%
and $g=\left\vert \det \left( g_{\mu \nu }\right) \right\vert $. IGE is the
asymptotic limit of the natural logarithm of the statistical weight defined
on $\mathcal{M}_{s}$ and represents a measure of temporal complexity of
chaotic dynamical systems whose dynamics is underlined by a curved
statistical manifold.

\section{Applications of the IGAC}

As a first example, we apply our IGAC to study the dynamics of a system with 
$l$ degrees of freedom, each one described by two pieces of relevant
information, its mean expected value and its variance (Gaussian statistical
macrostates). The line element $ds^{2}=g_{\mu \nu }\left( \Theta \right)
d\Theta ^{\mu }d\Theta ^{\nu }$ on $\mathcal{M}_{s}$ with $\mu $, $\nu =1$%
,..., $2l$ is defined by \cite{cafaro3, cafaro5},%
\begin{equation}
ds_{\text{Gaussians}}^{2}=\dsum\limits_{k=1}^{l}\left( \frac{1}{\sigma
_{k}^{2}}d\mu _{k}^{2}+\frac{2}{\sigma _{k}^{2}}d\sigma _{k}^{2}\right) 
\text{.}
\end{equation}%
This leads to consider an ED model on a non-maximally symmetric $2l$%
-dimensional statistical manifold $\mathcal{M}_{s}$. It is shown that $%
\mathcal{M}_{s}$ possesses a constant negative Ricci curvature that is
proportional to the number of degrees of freedom of the system, $\mathcal{R}%
_{\mathcal{M}_{s}}=-l$. It is shown that the system explores statistical
volume elements on $\mathcal{M}_{s}$ at an exponential rate. The information
geometrodynamical entropy $\mathcal{S}_{\mathcal{M}_{s}}$\ increases
linearly in time (statistical evolution parameter) and is moreover,
proportional to the number of degrees of freedom of the system, $\mathcal{S}%
_{\mathcal{M}_{s}}$ $\overset{\tau \rightarrow \infty }{\sim }l\lambda \tau $%
. The parameter $\lambda $ characterizes the family of probability
distributions on $\mathcal{M}_{s}$. The asymptotic linear
information-geometrodynamical entropy growth may be considered the
information-geometric analogue of the von Neumann entropy growth introduced
by Zurek-Paz, a \textit{quantum} feature of chaos. The geodesics on $%
\mathcal{M}_{s}$ are hyperbolic trajectories. Using the JLC-equation, we
show that the Jacobi vector field intensity $J_{\mathcal{M}_{s}}$ diverges
exponentially and is proportional to the number of degrees of freedom of the
system, $J_{\mathcal{M}_{s}}$ $\overset{\tau \rightarrow \infty }{\sim }%
l\exp \left( \lambda \tau \right) $. The exponential divergence of the
Jacobi vector field intensity $J_{\mathcal{M}_{s}}$ is a \textit{classical}
feature of chaos. Therefore, we conclude \ that $\mathcal{R}_{\mathcal{M}%
_{s}}=-l$, $J_{\mathcal{M}_{s}}\overset{\tau \rightarrow \infty }{\sim }%
l\exp \left( \lambda \tau \right) $, $\mathcal{S}_{\mathcal{M}_{s}}\overset{%
\tau \rightarrow \infty }{\sim }l\lambda \tau $. Thus, $\mathcal{R}_{%
\mathcal{M}_{s}}$, $\mathcal{S}_{\mathcal{M}_{s}}$ and $J_{\mathcal{M}_{s}}$
behave as proper indicators of chaoticity.

In our second example, we employ ED and "\textit{Newtonian Entropic Dynamics}%
" (NED) \cite{cafaro4}. In our special application, we consider a manifold
with a line element $ds^{2}=g_{\mu \nu }\left( \Theta \right) d\Theta ^{\mu
}d\Theta ^{\nu }$ (with $\mu $, $\nu =1$,..., $l$) given by \cite{cafaro7,
cafaro8},%
\begin{equation}
ds_{\text{IHOs}}^{2}=\left[ 1-\Phi \left( \Theta \right) \right] \delta
_{\mu \nu }d\Theta ^{\mu }d\Theta ^{\nu }\text{, }\Phi \left( \Theta \right)
=\overset{l}{\underset{k=1}{\sum }}u_{k}\left( \theta _{k}\right) \text{ }
\end{equation}%
where $u_{k}\left( \theta _{k}\right) =-\frac{1}{2}\omega _{k}^{2}\theta
_{k}^{2}$ and $\theta _{k}=\theta _{k}\left( s\right) $. The geodesic
equations for the macrovariables $\theta _{k}\left( s\right) $ are strongly 
\textit{nonlinear }and their integration is not trivial. However, upon a
suitable change of the affine parameter $s$ used in the geodesic equations,
we may simplify the differential equations for the macroscopic variables
parametrizing points on the manifold $\mathcal{M}_{s}$ with metric tensor $%
g_{\mu \nu }$. Recalling that the notion of chaos is observer-dependent and
upon changing the affine parameter from $s$ to $\tau $ in such a way that $%
ds^{2}=2\left( 1-\Phi \right) ^{2}d\tau ^{2}$, we obtain new geodesic
equations describing a set of macroscopic inverted harmonic oscillators
(IHOs). In order to ensure the compactification of the parameter space of
the system, we choose a Gaussian distributed frequency spectrum for the
IHOs. Thus, with this choice, the folding mechanism required for true chaos
is restored in a statistical (averaging over $\omega $ and $\tau $) sense.
Upon integrating these differential equations, we obtain the expression for
the asymptotic behavior of the IGE $\mathcal{S}_{\mathcal{M}_{s}}$, namely $%
\mathcal{S}_{\mathcal{M}_{s}}\left( \tau \right) \overset{\tau \rightarrow
\infty }{\sim }\Lambda \tau $ with $\Lambda =\overset{l}{\underset{i=1}{\sum 
}}\omega _{i}$. This result may be considered the information-geometric
analogue of the Zurek-Paz model used to investigate the implications of
decoherence for quantum chaos. They considered a chaotic system, a single
unstable harmonic oscillator characterized by a potential $V\left( x\right)
=-\frac{\Omega ^{2}x^{2}}{2}$ ($\Omega $ is the Lyapunov exponent), coupled
to an external environment. In the \textit{reversible classical limit }\cite%
{zurek2}, the von Neumann entropy of such a system increases linearly at a
rate determined by the Lyapunov exponent, $\mathcal{S}_{\text{quantum}%
}^{\left( \text{chaotic}\right) }\left( \tau \right) \overset{\tau
\rightarrow \infty }{\sim }\Omega \tau $.

In our final example, we use our IGAC to study the entropic dynamics on
curved statistical manifolds induced by classical probability distributions
of common use in the study of regular and chaotic quantum energy level
statistics. It is known \cite{prosen1, prosen2} that integrable and chaotic
quantum antiferromagnetic Ising chains are characterized by asymptotic
logarithmic and linear growths of their operator space entanglement
entropies, respectively. In this last example, we consider the
information-geometrodynamics of a Poisson distribution coupled to an
Exponential bath (spin chain in a \textit{transverse} magnetic field,
regular case) and that of a Wigner-Dyson distribution coupled to a Gaussian
bath (spin chain in a \textit{tilted} magnetic field, chaotic case).
Remarkably, we show that in the former case the IGE exhibits asymptotic
logarithmic growth while in the latter case the IGE exhibits asymptotic
linear growth. In the regular case, the line element $ds_{\text{integrable}%
}^{2}$ on the statistical manifold $\mathcal{M}_{s}$ is given by \cite%
{cafaro6, cafaro8},%
\begin{equation}
ds_{\text{integrable}}^{2}=\frac{1}{\mu _{A}^{2}}d\mu _{A}^{2}+\frac{1}{\mu
_{B}^{2}}d\mu _{B}^{2}
\end{equation}%
where the macrovariable $\mu _{A}$ is the average spacing of the energy
levels and $\mu _{B}$ is the average intensity of the magnetic energy
arising from the interaction of the \textit{transverse} magnetic field with
the spin $\frac{1}{2}$\ particle magnetic moment. In such a case, we show
that the asymptotic behavior of $\ \mathcal{S}_{\mathcal{M}_{s}}^{\left( 
\text{integrable}\right) }$ is sub-linear in $\tau $ (logarithmic IGE
growth), $\mathcal{S}_{\mathcal{M}_{s}}^{\left( \text{integrable}\right)
}\left( \tau \right) \overset{\tau \rightarrow \infty }{\sim }\log \tau $.
Finally, in the chaotic case, the line element $ds_{\text{chaotic}}^{2}$ on
the statistical manifold $\mathcal{M}_{s}$ is given by \cite{cafaro6,
cafaro8}, 
\begin{equation}
ds_{\text{chaotic}}^{2}=\frac{4}{\mu _{A}^{\prime 2}}d\mu _{A}^{\prime 2}+%
\frac{1}{\sigma _{B}^{\prime 2}}d\mu _{B}^{\prime 2}+\frac{2}{\sigma
_{B}^{\prime 2}}d\sigma _{B}^{\prime 2}
\end{equation}%
where the (nonvanishing) macrovariable $\mu _{A}^{\prime }$ is the average
spacing of the energy levels, $\mu _{B\text{ }}^{\prime }$and $\sigma
_{B}^{\prime }$ are the average intensity and variance, respectively of the
magnetic energy arising from the interaction of the \textit{tilted} magnetic
field with the spin $\frac{1}{2}$\ particle magnetic moment. In this case,
we show that asymptotic behavior of $\ \mathcal{S}_{\mathcal{M}_{s}}^{\left( 
\text{chaotic}\right) }$ is linear in $\tau $ (linear IGE growth), $\mathcal{%
S}_{\mathcal{M}_{s}}^{\left( \text{chaotic}\right) }\left( \tau \right) 
\overset{\tau \rightarrow \infty }{\sim }\tau $. The equations for $\mathcal{%
S}_{\mathcal{M}_{s}}^{\left( \text{integrable}\right) }$ and $\mathcal{S}_{%
\mathcal{M}_{s}}^{\left( \text{chaotic}\right) }$ may be considered as the
information-geometric analogues of the entanglement entropies defined in
standard quantum information theory in the regular and chaotic cases,
respectively.

\section{Conclusion}

In this paper we presented a novel theoretical information-geometric
framework suitable to characterize chaotic dynamical behavior of arbitrary
complex systems on curved statistical manifolds. Specifically, an
information-geometric analogue of the Zurek-Paz quantum chaos criterion of
linear entropy growth and an information-geometric characterization of
chaotic (integrable) energy level statistics of a quantum antiferromagnetic
Ising spin chain in a tilted (transverse) external magnetic field were
presented.

The descriptions of a classical chaotic system of arbitrary interacting
degrees of freedom, deviations from Gaussianity and chaoticity arising from
fluctuations of positively curved statistical manifolds are currently under
investigation. Furthermore, we are investigating the possibility to extend
the IGAC to quantum Hilbert spaces constructed from classical curved
statistical manifolds and we are considering the information-geometric
macroscopic versions of the Henon-Heiles and Fermi-Pasta-Ulam $\beta $%
-models to study chaotic geodesic flows on statistical manifolds. Finally,
the information geometry of a chaotic spring pendulum and a periodically
perturbed spherical pendulum are currently under investigation. Soft chaos
regimes arising in chemical physics are being considered as well. Our
objective is to study transitions from order to chaos in a floppy molecule
using inference methods and information geometry.

At this stage of its development, IGAC remains an ambitious unifying
information-geometric theoretical construct for the study of chaotic
dynamics with several limitations and unsolved problems \cite{cafaro5,
cafaro6}. However, based on our recent findings, we believe it could provide
an interesting, innovative and potentially powerful way to study and
understand the very important and challenging problems of classical and
quantum chaos.\ Therefore, we believe our research program deserves further
investigation and developments.

\begin{acknowledgments}
The author is grateful to S. A. Ali, Ariel Caticha and Adom Giffin for very
useful comments. Thanks are extended to all MaxEnt 2008 participants in Sao
Paulo- Brazil, especially to Rafael M. Gutierrez for his sincere interest
and very important comments on our works.
\end{acknowledgments}

\end{document}